\documentclass[conference]{IEEEtran}
\IEEEoverridecommandlockouts

\usepackage{cite}
\usepackage{amsmath,amssymb,amsfonts}
\usepackage{algorithmic}
\usepackage{graphicx}
\usepackage{textcomp}
\usepackage{xcolor}
\def\BibTeX{{\rm B\kern-.05em{\sc i\kern-.025em b}\kern-.08em
    T\kern-.1667em\lower.7ex\hbox{E}\kern-.125emX}}
\usepackage{datetime}
\usepackage[hyphens]{url}
\usepackage{fancyhdr}
\usepackage[bookmarks=true,breaklinks=true,letterpaper=true,colorlinks,citecolor=blue,linkcolor=blue,urlcolor=blue]{hyperref}
\usepackage{babel}
\usepackage{caption}
\usepackage{subcaption}

\title{SambaNova SN40L: Scaling the AI Memory Wall
with Dataflow and Composition of Experts}

\begin{document}
\bstctlcite{IEEEexample:BSTcontrol}
\makeatletter 
\newcommand{\linebreakand}{%
  \end{@IEEEauthorhalign}
  \hfill\mbox{}\par
  \mbox{}\hfill\begin{@IEEEauthorhalign}
}
\makeatother 

\pagenumbering{arabic}
\author{\IEEEauthorblockN{Raghu Prabhakar}
\and
\IEEEauthorblockN{Ram Sivaramakrishnan}
\and
\IEEEauthorblockN{Darshan Gandhi}
\and
\IEEEauthorblockN{Yun Du}
\and
\IEEEauthorblockN{Mingran Wang}
\linebreakand
\IEEEauthorblockN{Xiangyu Song}
\and
\IEEEauthorblockN{Kejie Zhang}
\and
\IEEEauthorblockN{Tianren Gao}
\and
\IEEEauthorblockN{Angela Wang}
\and
\IEEEauthorblockN{Xiaoyan Li}
\linebreakand
\IEEEauthorblockN{Yongning Sheng}
\and
\IEEEauthorblockN{Joshua Brot}
\and
\IEEEauthorblockN{Denis Sokolov}
\and
\IEEEauthorblockN{Apurv Vivek}
\and
\IEEEauthorblockN{Calvin Leung}
\linebreakand
\IEEEauthorblockN{Arjun Sabnis}
\and
\IEEEauthorblockN{Jiayu Bai}
\and
\IEEEauthorblockN{Tuowen Zhao}
\and
\IEEEauthorblockN{Mark Gottscho}
\and
\IEEEauthorblockN{David Jackson}
\linebreakand
\IEEEauthorblockN{Mark Luttrell}
\and
\IEEEauthorblockN{Manish K. Shah}
\and
\IEEEauthorblockN{Zhengyu Chen}
\and
\IEEEauthorblockN{Kaizhao Liang}
\and
\IEEEauthorblockN{Swayambhoo Jain}
\linebreakand
\IEEEauthorblockN{Urmish Thakker}
\and
\IEEEauthorblockN{Dawei Huang}
\and
\IEEEauthorblockN{Sumti Jairath}
\and
\IEEEauthorblockN{Kevin J. Brown}
\and
\IEEEauthorblockN{Kunle Olukotun}
\linebreakand
\IEEEauthorblockN{\large\textbf{SambaNova Systems, Inc.}}
\IEEEauthorblockA{first.last@sambanova.ai}
}

\maketitle
\begin{abstract}
Monolithic large language models (LLMs) like GPT\babelhyphen{nobreak}4 have paved the way for modern generative AI applications. 
Training, serving, and maintaining monolithic LLMs at scale, however, remains prohibitively expensive and challenging. The disproportionate increase in compute-to-memory ratio of modern AI accelerators have created a memory wall, necessitating new methods to deploy AI. 
Recent research has shown that a composition of many smaller expert models, each with several orders of magnitude fewer parameters, can match or exceed the capabilities of monolithic LLMs. 
\textit{Composition of Experts (CoE)} is a modular approach that lowers the cost and complexity of training and serving. 
However, this approach presents two key challenges when using conventional hardware: (1) 
without fused operations, smaller models have lower operational intensity, which makes high utilization more challenging to achieve; and (2) hosting a large number of models can be either prohibitively expensive or slow when dynamically switching between them. 

In this paper, we describe how combining CoE, streaming dataflow, and a three-tier memory system scales the AI memory wall. We describe 
Samba-CoE, a CoE system with 150 experts and a trillion total parameters.
We deploy Samba-CoE on the 
SambaNova SN40L Reconfigurable Dataflow Unit (RDU) -- a commercial dataflow accelerator architecture that has been co-designed for enterprise inference and training applications. 
The chip introduces a new three-tier memory system with on-chip distributed SRAM, on-package HBM, and off-package DDR DRAM. 
A dedicated inter-RDU network enables scaling up and out over multiple sockets. We demonstrate speedups ranging from $2\times$ to $13\times$ on various benchmarks running on eight RDU sockets compared with an unfused baseline.
We show that for CoE inference deployments, the 8-socket RDU Node reduces machine footprint by up to
$19\times$, speeds up model switching time by $15\times$ to $31\times$, and achieves
an overall speedup of $3.7\times$ over a DGX H100 and $6.6\times$ over a DGX A100.
\end{abstract}

\section{Introduction}
\label{sec:intro}
Recent advances in the training and inference of large language models (LLMs) has taken the world by storm. 
State-of-the-art generative AI/ML applications like ChatGPT~\cite{chatgpt} and Gemini~\cite{Bard} are built on top of \textit{monolithic LLMs} that can have billions or trillions of parameters. 
They are trained with curated datasets that consist of trillions of tokens scraped from the web. 
However, training and serving a state-of-the-art monolithic LLM is both an extraordinarily expensive affair and a complex systems engineering challenge. 
Training requires building and operating a supercomputer composed of thousands of hosts, purpose-built networks, power and cooling infrastructure, and thousands of accelerators -- typically GPUs~\cite{a100, h100} or TPUs~\cite{tpuv1, tpuv2_v3, tpuv4i, tpuv4}.  
The prohibitive cost and expertise required to train and serve 100s of billions of parameters put state-of-the-art AI capabilities out of reach for many academic researchers and smaller organizations, especially when on-premise deployments are needed. For instance, compute costs to train OpenAI's GPT-4 is estimated to be \$78 million USD, and Google's Gemini Ultra to be \$191 million USD~\cite{stanford_hai_2024}.
Building and deploying large monolithic models may not be sustainable for hyperscalers~\cite{copilot_money_loser} or any organization needing capable AI models continuously trained and updated on their data~\cite{costly1, costly2, costly3}. Finally, systems that cater to monolothic models have scaled compute TFLOPs much faster than memory bandwidth and capacity, creating the memory wall~\cite{memory_wall} where the memory system can no longer feed the compute efficiently.

The ML research community has responded with ecosystems of much smaller, modular models that are just as capable, but are cheaper and easier to train and serve~\cite{raffel_talk, mukherjee2023orca, gunasekar2023textbooks, jiang2023mistral}.
Smaller models like the 
8B-parameter Llama 3.1~\cite{llama-herd}, Llama2~\cite{llama2}, and Mistral 7B~\cite{mistral} are often adequate. They might not match the performance of larger models over a \textit{general} suite of tasks, but smaller models can deliver superior accuracy on a narrower set of \textit{specialized} tasks for which they have been fine-tuned~\cite{deepseek, code_llama}. 
For example, Flan-T5-XL has only 3B parameters, but it surpasses the 175B-parameter GPT-3's MMLU score by nearly 10\%~\cite{chung2022scaling}. 
Such proof points have bolstered community activity in training smaller models by specializing base models to a domain, by fine-tuning base models to a specific task or group of tasks~\cite{sanh2022multitask, wang2022supernaturalinstructions}, and by distilling or compressing larger models into smaller models. 
Furthermore, \emph{compositions} of such smaller models have been shown to demonstrate emergent behavior that matches large monolithic models~\cite{urmish1, urmish2, urmish3, urmish4, urmish5}. 
They bring AI within reach to a broader community. 

\begin{figure}
     \begin{subfigure}[b]{0.44\linewidth}
        \centering
         \includegraphics[width=\textwidth]{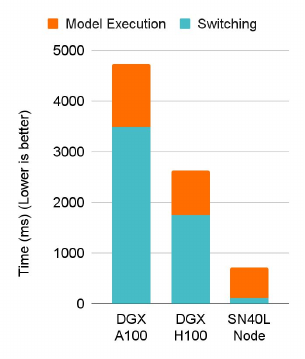}
         \caption{BS=8, TP=8}
         \label{fig:bs8-breakdown}
     \end{subfigure}
     \begin{subfigure}[b]{0.42\linewidth}
        \centering
         \includegraphics[width=\textwidth]{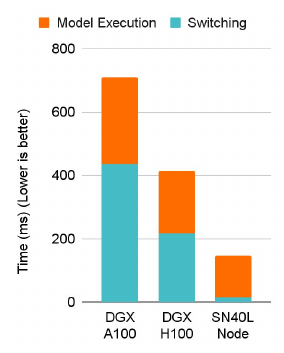}
         \caption{BS=1, TP=8}
         \label{fig:bs1-breakdown}
     \end{subfigure}
        \caption{CoE latency breakdown between model switching and model execution to generate 20 output tokens from a Llama2-7B expert. 
        The SN40L RDU executes CoEs efficiently by combining streaming dataflow and a novel three-tier memory hierarchy of SRAM, HBM, and DDR.
        }
        \label{fig:breakdown-intro}
\end{figure}
We believe that successful AI systems of the future will host and execute many small models efficiently. 
This is reflected both in directions pursued successfully in academia~\cite{raffel_talk, urmish2, urmish3, urmish4, urmish5}, and new products that are being adopted in industry~\cite{chai, samba1, samba-coe-v1}. CoE-like compound AI systems play a pivotal role in advancing the AI frontier~\cite{compound-ai-blog, multi-agent-blog}.
In this paper, we refer to such modular systems with compositions of specialized smaller models as \textbf{\emph{Composition of Experts (CoE)}}. 

A CoE consists of several small expert models working in tandem on a task. Outputs from one expert determine which expert(s) to execute next. Running an expert involves loading model parameter weights to the accelerator's main memory, and then executing the model. Consequently, executing a CoE involves a sequence of model switching and model execution. Current state-of-the-art AI accelerators do not handle this sequence of operations efficiently, as shown in Figure~\ref{fig:breakdown-intro}.

Efficiently accelerating a \textit{CoE} boils down to executing expert models efficiently while minimizing model switching costs. We break this down into three key requirements:
\begin{enumerate}
    \item \textbf{\emph{Aggressive} Operator Fusion and Pipeline Parallelism} to execute expert models efficiently. Smaller models have lower operational intensity~\cite{fusion_asplos22, s2d, williams_roofline_2009} and complex access patterns between operators ~\cite{flashfftconv}. Conventional operator fusion techniques~\cite{fusion-bert, fusion-tensorrt, pytorch2} achieve limited success across arbitrary access patterns.  
    \item \textbf{High-Bandwidth Memory} to exploit temporal and spatial locality in weights and intermediate results during generative inference, and
    \item \textbf{High-Capacity Memory} to minimize switching costs and store the parameters of many expert models.
\end{enumerate}

In this paper, we describe a hardware/software solution that overcomes the memory wall by addressing the challenges above.

We first describe the \textbf{Samba-CoE}, a trillion parameter CoE system with 150 8B expert models, and how running it efficiently requires hardware support for aggressive operator fusion and a novel memory system. 
We present the \textbf{SambaNova SN40L Reconfigurable Dataflow Unit (RDU)}, a commercial dataflow accelerator that combines \textbf{streaming dataflow parallelism} with a novel \textbf{three-tier memory system} containing large on-chip SRAM, HBM, and DDR DRAM that is directly attached to the accelerator.

The RDU's streaming dataflow architecture allows us to fuse \emph{hundreds} of complex operations with arbitrary access patterns into a single kernel call -- without requiring the programmer to write that kernel by hand.
This delivers large speedups by exploiting on-chip hardware support for mixtures of pipeline, data, and tensor parallelism. Our aggressive fusion techniques are well beyond the capabilities of state-of-the-art techniques used with conventional architectures~\cite{fusion_asplos22, fusion-tensorrt, fusion-bert, pytorch2}. 

Fabricated using TSMC 5nm technology, the SN40L RDU is a 2.5D Chip-on-Wafer-on-Substrate (CoWoS) chiplet-based design containing two SN40L Reconfigurable Dataflow Dies (RDDs) and HBM. 
Each SN40L RDU socket has 638 BF16 TFLOPS of peak compute performance using 1040 distributed Pattern Compute Units (PCUs). 
These are complemented by 1040 distributed Pattern Memory Units (PMUs) that in aggregate provide hundreds of TBps of on-chip memory bandwidth along with high bank-level parallelism within and across PMUs. Flexible on-chip address generation logic provides high bandwidth for arbitrary tensor memory access patterns.
The three memory tiers in SN40L are: 520 MiB of on-chip PMU SRAM, 64 GiB of co-packaged HBM, and up to 1.5 TiB of DDR DRAM (using pluggable DIMMs). 
Models are loaded from DDR to HBM at over 1 TB/s in a single SN40L Node.

We quantify and discuss the impact of streaming dataflow parallelism on several real world benchmarks, showing speedups ranging from \textbf{2$\times$} to \textbf{13$\times$} over an optimized baseline. We deploy Samba-CoE on a single \emph{SN40L Node} that contains eight SN40L RDU sockets and a host. We discuss the performance of Samba-CoE on the SN40L Node compared to DGX A100 and DGX H100. We show that for CoE inference deployments, the SN40L reduces machine footprint by up to \textbf{19$\times$}, speeds up model switching time by \textbf{15$\times$} to  \textbf{31$\times$}, and achieves an overall speedup of \textbf{3.7$\times$} to \textbf{6.6$\times$} over DGX H100 and DGX A100, respectively.

This paper is organized as follows: 
Section~\ref{sec:coe} describes Samba-CoE, our trillion parameter CoE. 
Section~\ref{sec:fusion} describes streaming dataflow and its challenges that translate to key hardware requirements.
Section~\ref{sec:rdu} describes the SN40L hardware architecture in detail and lists key changes from prior RDUs~\cite{sn10_hc, sn10_isscc}.
Section~\ref{sec:software} describes the software support for managing DDR and HBM. 
Section~\ref{sec:case_studies} quantifies the benefits of streaming dataflow as well as the performance of Samba-CoE. 
Section~\ref{sec:discussion} discusses key learnings from the hardware/software codesign process. Section~\ref{sec:related_work} covers related work.
We conclude in Section~\ref{sec:conclusion}.

\section{Background: Composition of Experts}
\label{sec:coe}
In this section, we describe one instance of a CoE built and deployed on the SN40L, called \textbf{\emph{Samba-CoE}}. Figure~\ref{fig:coe_1} shows the Samba-CoE pipeline from prompt to response. 

Samba-CoE consists of several expert models and a router model. Each expert is fine-tuned in a specific domain. We leveraged several excellent expert models fine-tuned on domains like coding, math, and language translation from the open source community. The router is another specialist model that dynamically assigns each input prompt to the most relevant expert. For instance, a math-related query would be routed to the math expert, while a coding question would go to the code expert.

The Samba-CoE is inspired by the Mixture-of-Experts (MoE) architecture~\cite{Jacobs_Jordan_Nowlan_Hinton_1991}, but has a few key differences. Although both MoEs and CoEs are more sparsely activated than a traditional dense monolithic model, MoEs are less flexible than CoEs. MoEs need to be trained/fine-tuned as a single model, similar to monolithic models, whereas CoEs are composed out of independently and heterogeneous expert models that are trained/fine-tuned independently of each other. CoEs are also more capable: prior work has shown that CoEs can outperform both MoEs~\cite{urmish2, urmish3} as well as large monolithic models like GPT-3.5 and GPT-4~\cite{samba1, samba-coe-v1}.
We note that CoEs and MoEs are orthogonal techniques that can be easily combined: a CoE can leverage expert models that are implemented internally as MoEs.

For simplicity, in this paper, the router model and expert models are all derived from Llama2-7B~\cite{llama2}. Note that the router and expert models do not need to be homogeneous - they can be different architectures with different numbers of parameters. Llama2-7B was chosen as the basis for this work due to its convenient size, impressive capabilities, and strong community support. The CoE concept and the Samba-CoE system are not limited to Llama2. 

\section{Hardware Requirements for CoE}
\label{sec:fusion}
\begin{figure}
    \centering
    \includegraphics[width=0.95\linewidth]{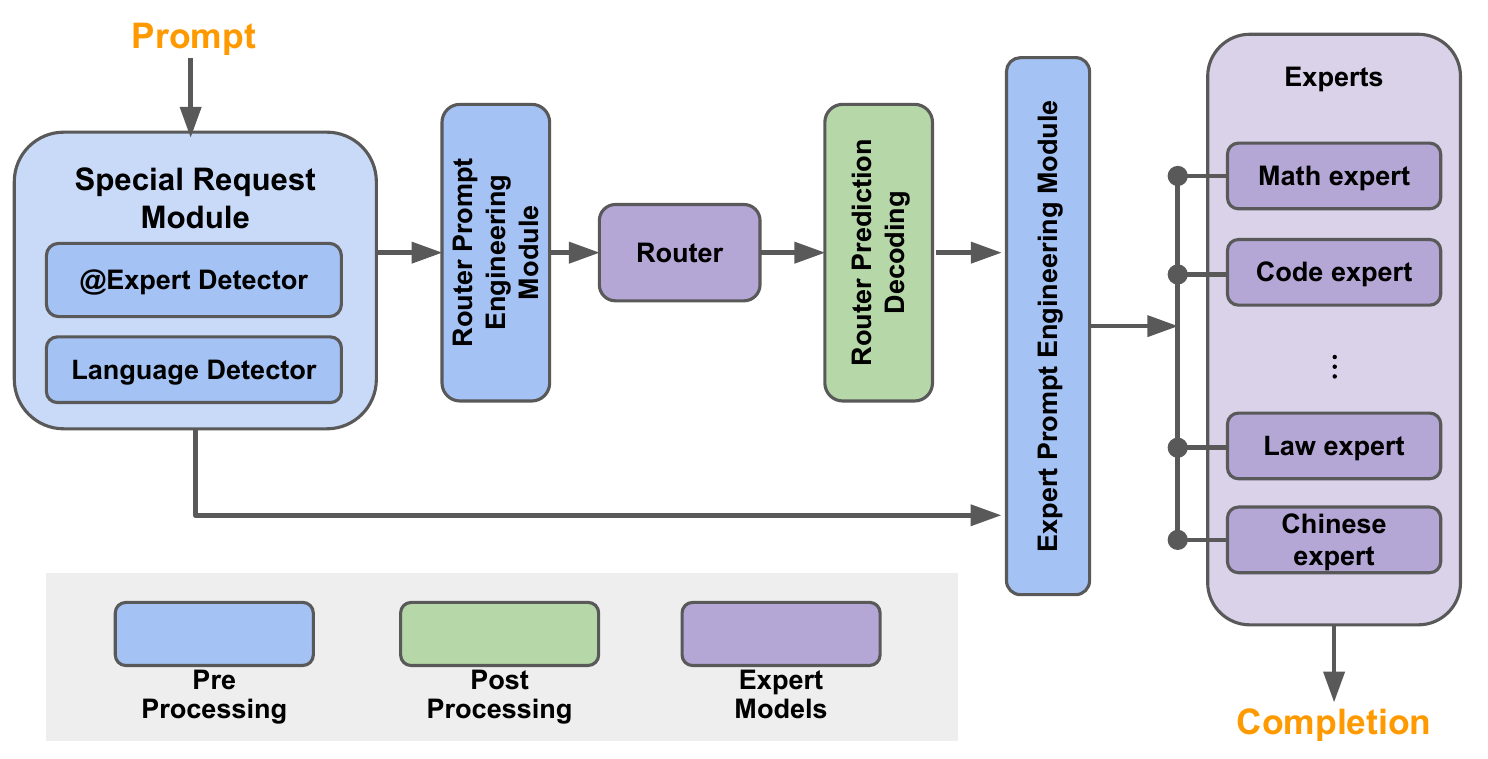}
    \caption{Samba-CoE Pipeline from prompt to completion.}
    \label{fig:coe_1}
\end{figure}

\begin{figure}
    \centering
    \includegraphics[width=0.95\linewidth]{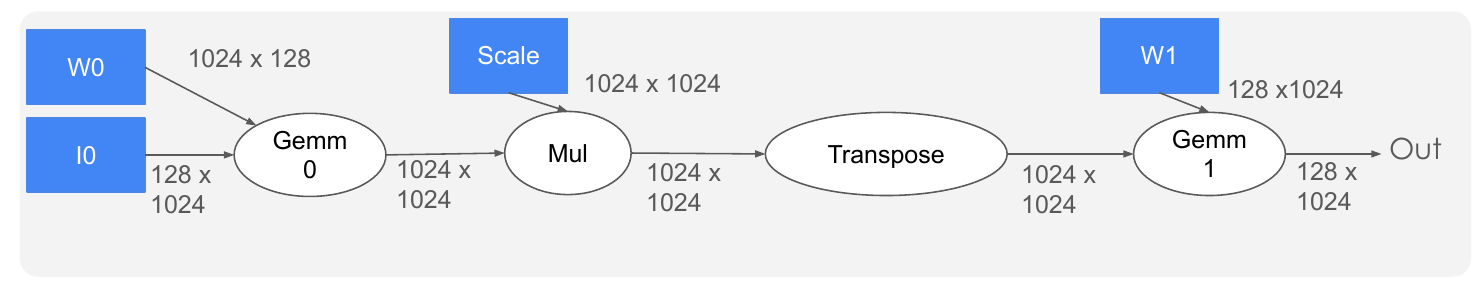}
    \caption{An example dataflow graph showing a simplified version of the Monarch FFT decomposition~\cite{flashfftconv}.
    }
    \label{fig:motivating}
\end{figure}
CoE execution time is broken down into model execution time and model switching time, as seen in Figure~\ref{fig:breakdown-intro}. Minimizing CoE execution time can be used to either reduce the machine footprint per user or increase the number of users supported under a given footprint. To reduce model execution time, we show the advantages of streaming dataflow over conventional operator fusion. To minimize model switching time, we motivate the need for both a high-capacity accelerator-local DDR interface and HBM.

\subsection{Streaming Dataflow}

\noindent \textbf{Conventional Operator Fusion is Insufficient:}
Operator fusion is a common optimization technique to increase operational intensity and improve hardware utilization~\cite{fusion_asplos22, flashattention, flashattention2, flashfftconv, fusion-tensorrt, fusion-bert, pytorch2}. 
Fusion also reduces the number of kernel calls required to run a model and amortizes kernel launch overheads.
However, expert models often contain operators with low operational intensity~\cite{mamba, striped_hyena} coupled with complex access patterns involving shuffles and transposes~\cite{flashfftconv}. Complex access patterns severely restricts the efficacy of fusion on GPUs. Frameworks like PyTorch2~\cite{pytorch2} and TensorRT~\cite{fusion-tensorrt} have documented restrictions on patterns that are explicitly not supported for fusion. Consequently, many complex fused kernels are still handwritten~\cite{flashattention, flashattention2} for GPUs.

Figure~\ref{fig:motivating} depicts a simplified Monarch FFT decomposition~\cite{flashfftconv} with tensor shapes annotated on the edges.
Table~\ref{tab:motivation_op_intensity} shows the impact of fusion on the operational intensity. 
Higher operation intensity allows applications to achieve roofline performance for a given target accelerator. For instance, an A100 GPU has a TFLOPS/TBps ratio of approximately $300 / 2 = 150$, meaning kernels with operation intensities less than 150 FLOPs/byte are memory-bound on the A100.
In Table~\ref{tab:motivation_op_intensity}, the first two rows are memory bound on A100, and the last row is compute bound. 
\begin{table}
    \centering
    \begin{tabular}{|c|c|} \hline 
         \textbf{Fusion Level} & \textbf{Operation Intensity (Ops / Byte)}\\ \hline 
         No Fusion& 39.5\\ \hline 
         Gemm0 - Mul - Transpose & 102.6\\ \hline 
         Fully Spatially Fused & 410.4\\ \hline
    \end{tabular}
    \vspace{5pt}
    \caption{Impact of different levels of fusion on operation intensity for the example in Figure~\ref{fig:motivating}. Without full fusion, this example will be memory bound on most architectures.}
    \label{tab:motivation_op_intensity}
\end{table}

However, GPUs cannot fuse all of Figure~\ref{fig:motivating} for the following reasons:
\begin{enumerate}
    \item Rigid memory hierarchy and programming model creates data movement bottlenecks: A GPU kernel is launched with a grid of thread blocks. The grid structure is fixed for the duration of the kernel. Fusing \textit{Gemm0} and \textit{Mul} would be trivial. However, \textit{Transpose} forces threads to access values from threads in other SMs, triggering a data exchange across SMs via the shared cache and HBM. As there is no other means to transfer data between SMs, this lack of flexibility creates a bottleneck at the shared cache and HBM 
    \item Insufficient on-chip SRAM capacity 
    forces materialization of the output of \textit{Transpose} to HBM, preventing a fusion opportunity.
    \item No pipeline parallelism exploited between operators: Higher order Monarch FFT decompositions (studied in Section~\ref{sec:case_studies}) create many small matrix multiplies that are $32\times32\times32$ or smaller, which do not utilize all SMs efficiently. However, there is abundant pipeline-level parallelism across all the matrix multiplies and element-wise operators. The GPU SIMT programming model does not provide a straightforward way to execute the operators in Figure~\ref{fig:motivating} as a pipeline.  
\end{enumerate}

\noindent \textbf{Streaming Dataflow enables Pipelining and Automatic Fusion with Arbitrary Access Patterns:}

Unlike conventional fusion, \emph{streaming dataflow} executes operators as a coarse-grained pipeline. Tensors are tiled and streamed through this pipeline. Tiles can have \emph{any arbitrary read and write access patterns} between operators.
 
Figure~\ref{fig:motivating-df} depicts the spatially fused implementation. Blue boxes represent on-chip buffer units, and gray boxes represent on-chip compute units. The operators \texttt{Gemm0}, \texttt{Mul}, and \texttt{Gemm1} are executed as stages in a coarse-grained pipeline. The blue memory units in between serve as decoupling stage buffers that hold intermediate results. More compute units are assigned to \texttt{Gemm0} and \texttt{Gemm1} as they account for a larger fraction of the total operations. Input and output bandwidths to and from stage buffers are matched to their respective stages by using the appropriate number of memory units. For instance, logical stage buffer \texttt{I0} is partitioned into two memory units \texttt{I00} and \texttt{I01} to match the required input bandwidth to \texttt{Gemm0}. Buffer \texttt{S0} - \texttt{S3} is partitioned into four memory units for capacity reasons.  The transpose operation is fused into buffers \texttt{T0*} and \texttt{T1*} as an access pattern. 

We distill the observations from above into the following list of on-chip architecture features to enable streaming dataflow:
\begin{enumerate}
    \item \textbf{Composable memory units}: A single memory unit provides a fixed capacity and bandwidth. As capacity and bandwidth needs vary across on-chip tensors, hardware should support programmable interleaving of logical addresses across memory units.
    \item \textbf{Address generation bandwidth and flexibility}: High data bandwidth requires high address generation bandwidth. 
    Furthermore, each memory unit should support non-blocking concurrent reads and writes to implement stage buffers efficiently. In short, the address generation hardware should allow generating multiple concurrent addresses at high throughput for arbitrarily complex address expressions.
    \item \textbf{Systolic and streaming compute}: ML accelerator architectures often implement systolic arrays to increase compute density for GEMM-like operations. However, in many ML models, GEMMs are frequently followed by element-wise operators and reductions which require high throughput streaming compute capability. 
    \item  \textbf{One-to-many, many-to-one, and data reordering}: Disparities between the number of producer and consumer units create one-producer-to-many-consumers and many-producers-to-one-consumer traffic streams that also require flow control. For one-to-many, hardware support is required to create fan-out paths in the interconnect from the source to a program-decided set of destinations. For many-to-one traffic, data from different paths can arrive out-of-order at the destination. The out-of-order sequence must often be put back in order to match the program's expectation at the destination. In other words, hardware must provide a protocol to reorder data streams. Finally, program-controlled bandwidth management (e.g., throttling) and routing are necessary to satisfy streams with differing bandwidth requirements. All of this hardware support needs to be usable by automatic place-and-route algorithms within the compiler.
\end{enumerate}
\begin{figure}
    \centering
    \includegraphics[width=0.95\linewidth]{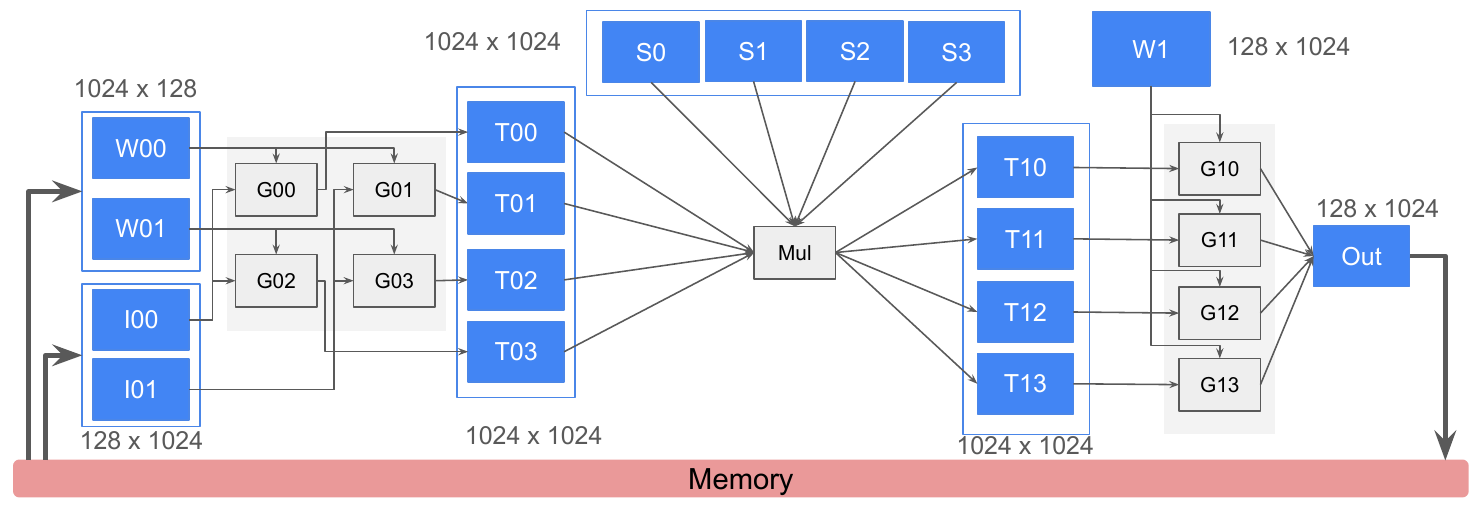}
    \caption{A spatially fused implementation of Figure~\ref{fig:motivating}. Blue boxes are on-chip buffers, gray boxes are compute units, and edges are on-chip communication.
    }
    \label{fig:motivating-df}
\end{figure}

\subsection{Model Hosting and Switching Costs} 
HBM's limited capacity limits the number of experts that can be in a CoE when hosted on a GPU or TPU. With HBM alone, running large CoEs requires either (a) using more machines for HBM capacity, which increases costs, complicates deployment, and introduces load balancing challenges, or (b) using the host's memory, which increases switching latency, as shown in Figure~\ref{fig:breakdown-intro}. 
Higher capacity DDR memory that is attached directly to the accelerator reduces both model hosting and model switching costs. Furthermore, CoEs exhibit temporal locality in expert parameters, as they are used multiple times (during autoregressive decoding, for instance). HBM plays a key role in exploiting this temporal data locality by acting as a software-managed caching tier between DDR and SRAM.

Consequently, we conclude that systems to execute composition of smaller models need two types of off-chip memories: (1) \textbf{high-bandwidth} memory to exploit temporal locality of expert parameters, and (2) \textbf{high-capacity} memory to store expert parameters in a small footprint.

In the next section, we describe the SN40L Reconfigurable Dataflow Unit which is built on the above principles.
\section{SN40L Hardware Architecture}
\label{sec:rdu}
The SN40L dataflow accelerator is fabricated using TSMC's 5FF process and packaged as a dual die socket using Chip-on-Wafer-on-Substrate (CoWoS) multi-chip packaging technology. Table~\ref{tab:sn40l} lists key chip parameters for the SN40L RDU.
Figure~\ref{fig:sn40-block-diagram} shows the salient components of the SN40L, which are described below.

    \noindent \textbf{RDU Tile}: A coarse-grained reconfigurable array of dataflow cores. Consists of Pattern Compute Units (PCUs), Pattern Memory Units (PMUs),
    and
    Address Generation and Coalescing Units (AGCUs)
    that are connected together in a two-dimensional mesh interconnect called the Reconfigurable Dataflow Network (RDN). 
    
    \noindent \textbf{Memory Interfaces}: The SN40L interfaces with two tiers of off-chip memories -- HBM and DDR. Both memory spaces are software managed. The DDR tier can have a peak memory capacity of 1.5 TiB at a peak bandwidth of over 200 GB/s. The HBM tier has 64 GiB of capacity with a peak bandwidth of about 2 TB/s per socket.
    
    \noindent \textbf{Die-to-Die (D2D) Interface}: SN40L tile components can stream data between two dies directly without going through off-chip memory. 
    
    \noindent \textbf{Host Interface}: SN40L interfaces with a host x86 CPU using a PCIe link. This interface supports DMA between host and device off-chip memory as well as direct communication between the host and the tile.  
    
    \noindent \textbf{Peer-to-Peer (P2P) Interfaces}: Connects an SN40L to other SN40L RDUs. A peer-to-peer protocol described in section~\ref{subsec:agcu} provides primitives to implement collective communication primitives.
    
    \noindent \textbf{Top Level Network (TLN)}: This network connects an SN40L tile to the host, memory, and peer-to-peer interfaces.

\begin{figure}
    \centering
    \includegraphics[width=0.7\linewidth]{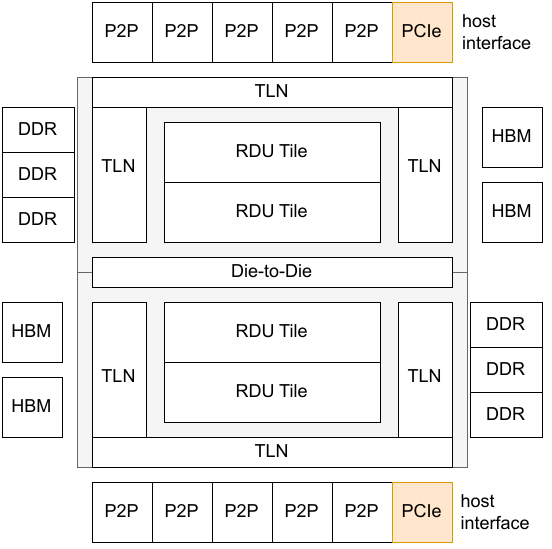}
    \caption{Block Diagram of the 2-die SN40L showing high-level components and interfaces.
    }
    \label{fig:sn40-block-diagram}
\end{figure}

\begin{table}
\centering
\def\arraystretch{1.5}%
\begin{tabular}{|p{1in} |p{1.5in}|} \hline
\textbf{Parameter} & \textbf{Value} \\ \hline
Compute Capability     & 638 BFLOAT16 (BF16) TFLOPs \\ \hline
SRAM Capacity     & 520 MB \\ \hline
HBM Capacity     & 64 GB \\ \hline
HBM Bandwidth     & 1.8 TB/s \\ \hline
DDR Capacity     & 1.5 TB \\ \hline
DDR Bandwidth     & 200 GB/s \\ \hline
PCU Count       & 1040 \\ \hline
PMU Count       & 1040 \\ \hline
Clock Frequency       & $<$ 2 GHz \\ \hline
Technology, Die Size  & 5nm, $<$ 650 mm\textsuperscript{2} \\ \hline
Dies per socket  & 2 \\ \hline
\end{tabular}
\caption{Chip parameters for the SN40L RDU.}
\label{tab:sn40l}
\end{table}

Figure \ref{fig:rdu-tile} illustrates an SN40L tile with the key dataflow components: PCUs, PMUs, RDN switches, and AGCUs. The following subsections describe them in more detail. 
\begin{figure}
    \centering
    \includegraphics[width=1\linewidth]{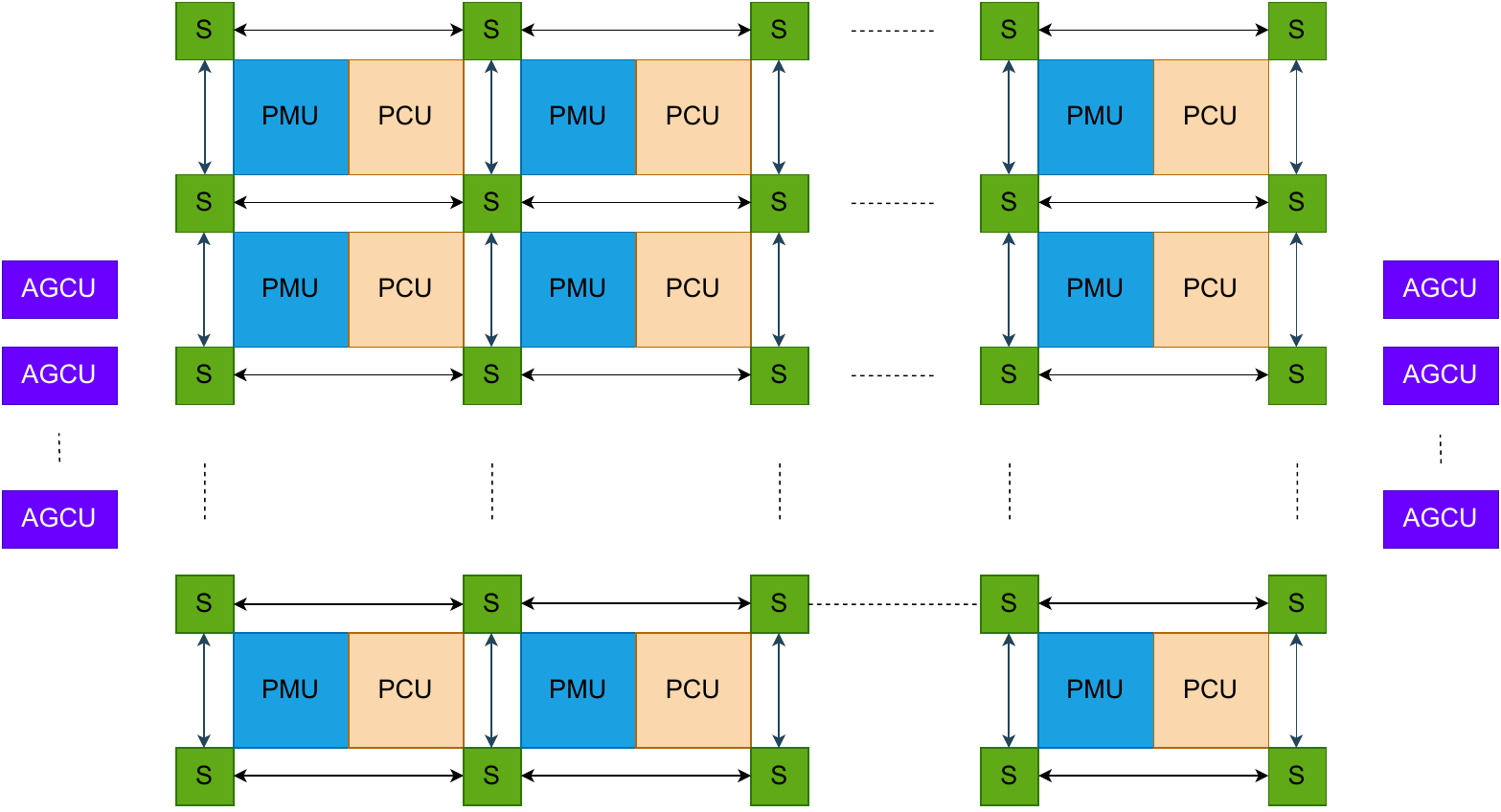}
    \caption{Block Diagram of the SN40L Tile: PCUs, PMUs, RDN switches, AGCUs. Connections between AGCU and switches is not shown. }
    \label{fig:rdu-tile}
\end{figure}

\subsection{Pattern Compute Unit (PCU)}
\label{subsec:PCU}
\begin{figure}
    \centering
    \includegraphics[width=0.95\linewidth]{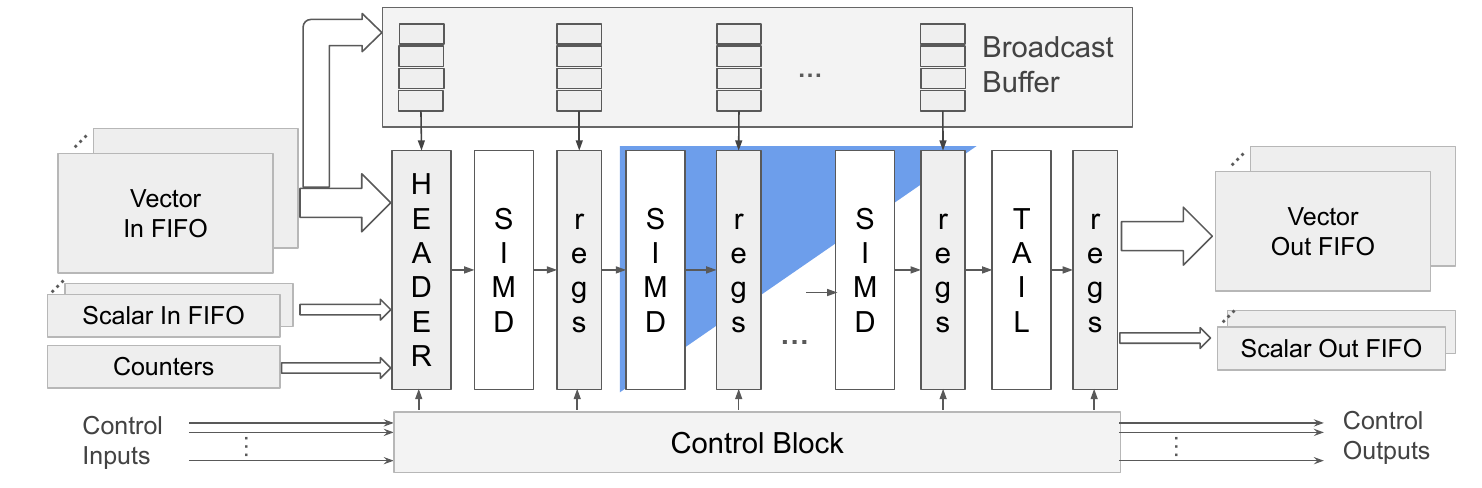}
    \caption{PCU block diagram showing components for systolic and SIMD operation with cross-lane reduction.}
    \label{fig:PCU-systolic-streaming}
\end{figure}

The PCU in SN40L provides the systolic and streaming compute capabilities in the SN40L. The PCU's datapath consists of a header, body, and tail. 
The header consumes incoming dataflows and drives the body section.
The PCU's body section is configurable as an 
output stationary systolic array or as a 
pipelined SIMD core with  
multiple stages of vector compute.
The tail performs special element-wise functions and drives the output FIFOs.
This enables efficient execution of GEMM-like operations, element-wise operations, or reductions.

Figure~\ref{fig:PCU-systolic-streaming} illustrates the PCU as both 2D systolic array and as a SIMD core. The systolic array accelerates matrix multiplications like \emph{Gemm0} and \emph{Gemm1} in Figure~\ref{fig:motivating}. 
Inputs to the systolic array are streamed left-to-right and top-to-bottom through a structure called \emph{broadcast buffer}. Accumulated results are drained left-to-right to output FIFOs through the tail unit. 
Matrix multiplication can be parallelized further across multiple PCUs, similar to the depiction in Figure~\ref{fig:motivating-df}. 
As a SIMD core, the PCU executes a parallel multidimensional tensor operation in a fully pipelined fashion, like \emph{Mul} in Figure~\ref{fig:motivating-df}. Each SIMD stage supports common arithmetic, logical, and bit-wise operations in FP32, BF16, and INT32 formats. 
The PCU can be configured to implement an optional cross-lane reduction network, shown as the blue triangle in the figure. Lane-wise reductions are also supported, in pure SIMD fashion.
Counters track loop iterations and generate control events when they reach the programmed maximum value, indicating that a loop has completed execution.

The tail section supports transcendental functions, random number generation, stochastic rounding, and format conversions. A tail operation can be fused and pipelined with compute in the body section. 

An operation can be parallelized across multiple PCUs in a data parallel, tensor parallel, or pipeline parallel fashion. Data parallelism is achieved by partitioning inputs and outputs to create multiple independent data streams that are processed by different PCUs. 
Tensor parallelism is achieved by forking into data parallel streams, then joining them. 
Pipeline parallelism is achieved by chaining multiple PCUs together to fuse operations and increase operational intensity.

\subsection{Pattern Memory Unit (PMU)}
\label{subsec:PMU}
\begin{figure}
    \centering
    \includegraphics[width=0.95\linewidth]{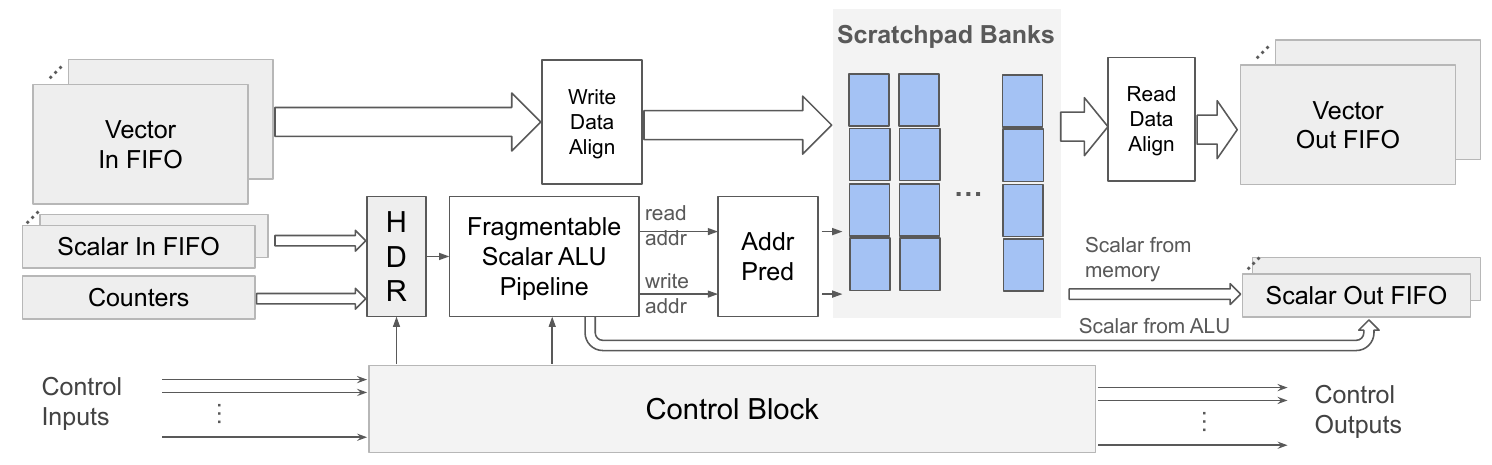}
    \caption{PMU ALUs, predication, and data alignment units.
    }
    \label{fig:PMU}
\end{figure}
The PMU in SN40L provides the on-chip memory capacity, bandwidth, and addressing flexibility for efficient operator fusion. Figure~\ref{fig:PMU} shows the high-level block diagram of the PMU. PMUs are used to store on-chip tensors like inputs, parameters, metadata, and intermediate results. For example, all blue blocks in Figure~\ref{fig:motivating-df} correspond to PMUs. The key PMU components are described below.

    \noindent \textbf{Scratchpad banks:} Each PMU contains a programmer-managed 
    scratchpad memory that is organized as an array of 
    SRAM instances. The SRAM array collectively supports concurrent writes and reads.
    
    \noindent \textbf{Scalar ALU Pipeline:} A PMU contains 
    several stages of integer 
    ALUs that can be configured to generate read and write addresses concurrently to flexibly access a tensor in the scratchpad.  
    PMU ALUs implement a set of special complex instructions like bitfield extraction and shift-and-set, that are frequently used in address computations. This enables producing complex addresses more efficiently with fewer ALU stages, and hence lesser latency. The ALU pipeline also has a path to ingest scalars as operands from the scalar RDN, and output computed values as scalars into the scalar RDN. This path facilitates addressing composability: complex integer calculations can be broken up and mapped across several PMUs if needed. 

    As described in section~\ref{sec:fusion} and shown in Figure~\ref{fig:motivating-df} (buffers \texttt{T00-T03} and \texttt{T10-T13}), stage buffers in a spatially fused kernel require concurrent reads and writes which may have different access patterns. While not universally true, we have anecdotally observed scenarios where write and read access patterns for a tensor inversely affect each other's complexity; a complex write access pattern often enables a simpler read access pattern and vice versa. The scalar ALU pipeline allows software to exploit this insight. It can be partitioned into independent read and write address generation pipelines with a software-configured number of stages allocated to each access.
    
    \noindent \textbf{Address Predication and Banking:} Figure~\ref{fig:motivating-df} shows that a single logical tensor can span multiple PMUs due to capacity needs (\texttt{S0-S3}), bandwidth needs (\texttt{W00-W01} and \texttt{I00-I01}), or both (\texttt{T00 - T03} and \texttt{T10 - T13}). The PMU enables this by providing hooks to programmatically control tensor address interleaving across PMUs. Specifically, each PMU can be programmed with a range of valid addresses for that PMU, or a programmable predicate bit per generated address. An addresses is processed if it is within the programmed range or a valid predicate, and dropped otherwise. Furthermore, addresses are mapped to scratchpad banks using bank bit locations that can be programmed by software. 
    
    \noindent \textbf{Data Alignment Unit:} The data alignment unit supports common tensor transformation operations like transpose, cross-lane vector permute, vector-unaligned accesses, lookup table (LUT), data format, and data layout conversions. 
    Tensors to be transposed are written in a special diagonally striped format across the scratchpad banks that enables reading the same tensor in both regular and transposed format at full bandwidth. This capability enables implementing the \emph{transpose} operator in Figure~\ref{fig:motivating} as a read-write access pattern optimization between buffers \texttt{T00-T03} and \texttt{T10-T13} in Figure~\ref{fig:motivating-df}.
 
\subsection{Reconfigurable Dataflow Network (RDN)}
\label{subsec:RDN}
The RDN is the on-chip programmable interconnect that facilitates communication between PCUs, PMUs, and AGCUs. The RDN consists of three physical fabrics - vector, scalar, and control. The vector and scalar fabrics are packet-switched. The control fabric is circuit-switched and consists of a bundle of single bit wires that can be individually routed. 
The vector fabric is the primary conduit for tensor data. The scalar fabric is mainly used to transport metadata such as a addresses but in some cases it can also be used to carry data or control. The control fabric is used to carry tokens for distributed coarse-grain flow control and to collectively orchestrate the execution of a graph. Control tokens typically correspond to counter `done' events that indicate the end of a loop, as discussed briefly in Section~\ref{subsec:PCU}.

The RDN is implemented using a mesh of non-blocking switches.
As shown in Figures~\ref{fig:PCU-systolic-streaming} and ~\ref{fig:PMU}, inbound scalar and vector packets to units from switches land in input FIFOs, and exit via output FIFOs. Transmissions on the vector and scalar fabric are subject to credit-based flow control at every hop. Packet streams are also end-to-end flow controlled between communicating units on the RDN through a combination of coarse-grained software tokens, fine-grained hardware credits, and forward progress guarantees in hardware.
The routing tables for all three RDN fabrics are configured by software using a place-and-route (PnR) layer within the compiler.

We now describe the mechanics of supporting the key communication patterns identified in Section~\ref{sec:fusion}.
\begin{itemize}
    \item \textbf{Multi-cast and Programmable routing}: Routing of packets on the scalar and vector fabrics is done either dynamically using a 2-D dimension order route or as software-controlled static flow routes. In static flow routing, software assigns a flow ID field to a packet stream, which is carried with the packet. The flow ID field is decoded at every switch port and reassigned prior to forwarding the packet to its next destination. The static flow routing mechanism supports packet multi-casting through the switches.
    
    \item \textbf{Many-to-one and Data Reordering}: Vector packets contain a metadata field called \emph{sequence ID}, which is the primary mechanism to support arbitrary many-to-one streams on-chip. Vector output ports of units are equipped with programmable logic to generate sequence IDs for each output vector. This way, sequence IDs can be programmed by software to represent the logical vector order for a given operation across multiple sources. The sequence ID field is used as an input operand in a PMU to compute the write addresses to reorder the packets. 
\end{itemize}

\subsection{Address Generation and Coalescing Unit (AGCU)} 
\label{subsec:agcu}
The AGCU is a reconfigurable dataflow bridge for the RDU tile to access local device memory (HBM/DDR), host memory, remote RDU device memory, and remote RDU tiles via the TLN.
On the tile-side, it acts like a dataflow core by exposing RDN vector, scalar, and control ports.
On the TLN-side, it generates read and write requests and coalesces the responses.
It is equipped with a scalar address generation pipeline and counters, bearing some similarities to the PMU logic (sans SRAM).
It also provides an address translation layer for memory management.

\noindent \textbf{Peer-to-Peer:} The AGCU supports a peer-to-peer (P2P) communication protocol to directly stream data between RDU tiles on different sockets without involving DDR or HBM. The P2P protocol enables building collective communication primitives between RDUs such as \emph{AllReduce}.

\noindent \textbf{Kernel Launch Orchestration:} The AGCU implements a kernel launch mechanism which consists of sequence of three commands: Program Load, Argument Load, and Kernel Execute. Running a model involves executing a schedule of kernel launches, which can be \emph{software-orchestrated} or \emph{hardware-orchestrated}. Software orchestration allows more flexible scheduling of kernels and provides more host software visibility into model execution. However, software orchestration incurs overheads that can impact performance. Hardware orchestration offloads a static kernel schedule to the dedicated hardware in the AGCUs, which significantly reduces the overheads. In section~\ref{sec:case_studies}, we quantify and discuss the impact of software vs. hardware-orchestrated execution on various benchmarks.
\section{Software Support}
\label{sec:software}
Here we describe how Samba-CoE is deployed on SN40L.

\begin{figure}
    \centering
    \includegraphics[width=0.95\linewidth]{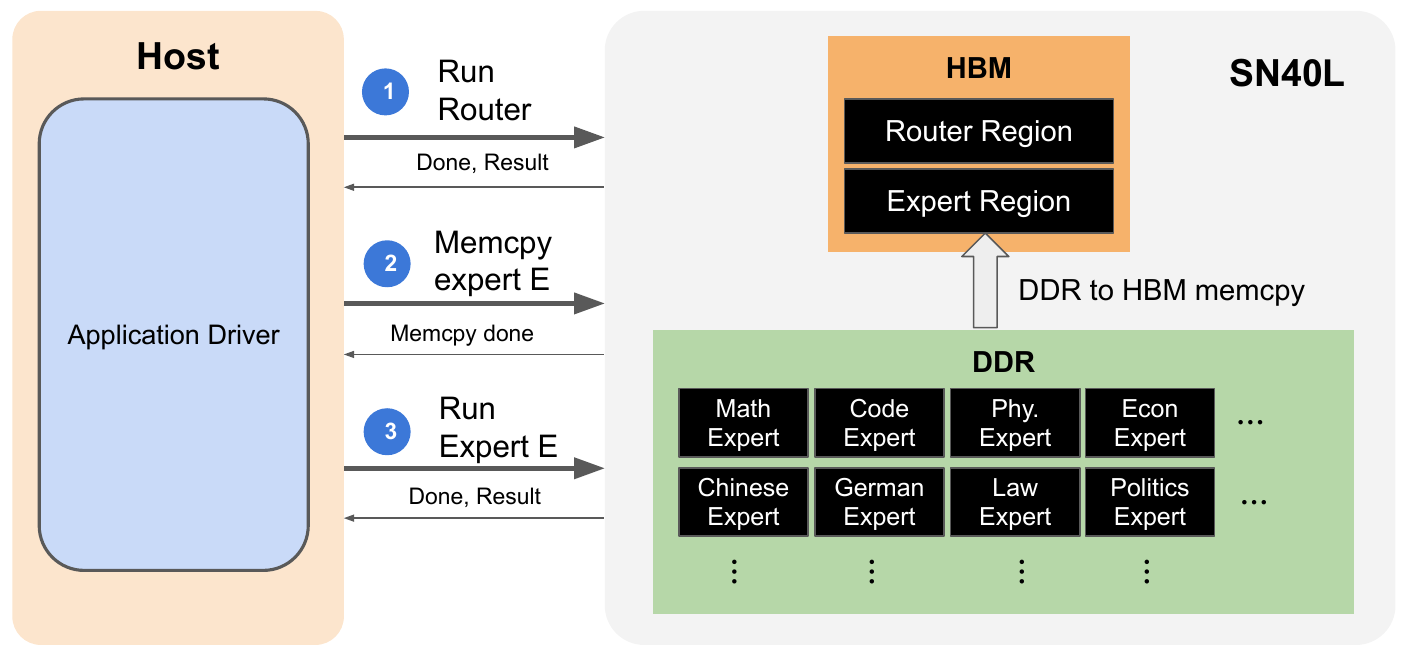}
    \caption{Simplified sequence of operations for Samba-CoE on SN40L. Router weights are in HBM. Expert weights are in DDR, with a region pre-allocated in HBM for the ``current" expert(s).}
    \label{fig:coe_2}
\end{figure}

Samba-CoE consists of 150 Llama2-7B experts with a total of over 1T parameters.
It is deployed on a single SN40L node with eight RDU sockets. Figure~\ref{fig:coe_2} shows how Samba-CoE leverages both DDR and HBM, along with a simplified flow of events for a single prompt. Weights for all 150 experts are held in high capacity DDR, while weights for the router are held in HBM. A single Samba-CoE inference has three high-level steps: (1) Run the router to determine the expert, (2) Copy the expert from DDR to HBM, and (3) Run the expert. After the model switch, the expert model's weights are read multiple times in the auto-regressive decoding loop to generate multiple tokens. This inherent model-level temporal locality in Samba-CoE is exploited by moving the weights to HBM.

However, this does impose additional burdens on the software stack to manage multiple non-uniform memory spaces. In this section, we discuss the software infrastructure we built in between the application layer and the SN40L's low-level runtime driver to implement this vision. 

\subsection{Memory Allocation}
One of our goals for this system design was to make device memory management transparent to the application developer. 
To that end, we added automatic heterogeneous device memory management into the SN40L compiler. The starting assumption is to use HBM by default as long as everything fits. We therefore use DDR for two main purposes: 1) spilling data from HBM to DDR when a given model's resident memory is too large to fit in HBM, and 2) holding all of the other models that are part of the CoE but are currently inactive. Note that while there is a nontrivial performance cost in moving data between HBM and DDR, it's still significantly cheaper than it would be to spill all the way back to host's memory, as quantified in Figure~\ref{fig:breakdown-intro}). 
In this section, we  focus on the first use case, and discuss the second in Section \ref{sec:runtime}.

We found that aggressive garbage collection is required to fit models like Llama2 7B from Table \ref{tab:benchmarks} in the 64 GiB HBM capacity per socket. However typical dynamic garbage collection schemes have far too much overhead for this use case, as it would require us to frequently return control back to the CPU to reorganize the SN40L's memory in the middle of the application. Instead we exploited the fact that the SN40L's programming model has neither dynamic memory allocation nor pointer aliasing, so we can therefore perform symbol lifetime analysis statically and implement garbage collection by assigning multiple logical symbols to the same device virtual addresses as long as their lifetimes don't overlap. 


Finally, symbols to be spilled to DDR are determined when the memory still doesn't fit. As there is an order of magnitude bandwidth difference between the two memories, we analyze the temporal locality of each symbol and its transfer footprint to estimate the total bandwidth requirement for that symbol over the entire application. The symbols are then sorted by their aggregate transfer size, so that we will spill symbols with the smallest bandwidth requirement first. In practice we've found that for the LLM models in Table \ref{tab:benchmarks}, the weights receive highest priority to remain in HBM, while activation symbols and other intermediate results can be spilled if necessary. We are investigating further improvements to this algorithm.

\subsection{CoE Runtime}
\label{sec:runtime}
One of our primary design goals for creating a CoE model architecture is that the lifecycle of each individual model expert should be independent: developing, compiling, training, fine-tuning, quantizing, maintaining, serving, and sharing. This provides a significant leap in software modularity over monolithic models. Progress in model accuracy and performance is rapid, and CoE makes it easy to incrementally leverage the latest innovations.

To achieve this goal, we need to link an arbitrary number of independently compiled models together at runtime and switch between them dynamically as new requests come in from the application layer. The approach is similar to how a dynamic linker/loader works in traditional software applications.

We added a lightweight dynamic memory manager on top of the static manager described previously. It can be efficiently implemented using the host runtime because it only has to run at the boundary of an entire expert model. Each compiled model binary tells us ahead of time exactly how much HBM and DDR space that model will require. The CoE runtime then interfaces with the low-level device driver to dynamically allocate blocks of memory in DDR for each model. This allocation includes the memory that the compiler intended for HBM; those memory segments are also stored in DDR initially. When a particular model is requested, the CoE runtime ``activates'' it by copying the memory segments intended for HBM, and then transfers control to the compiled application binary to run as it normally would. Once the application exits, control is returned to the CoE runtime which waits for the next request.

The CoE runtime tries to keep as many models as possible active in HBM at a time and uses an LRU eviction policy when we hit the capacity limit. If the next request is for the same model, it can resume immediately with no additional overhead. If the request requires a new model that won't fit in the remaining HBM space, then the oldest model's HBM memory evicted from HBM 
, and the new model's HBM memory is swapped in. 
To avoid copying back large read-only weights unnecessarily, the compiler annotates symbols as read-only where appropriate, and the runtime then skips copying any such symbols back to DDR.

\section{Case Studies}
\label{sec:case_studies}
In this section, we quantify and discuss the impact of operator fusion using several real world training and inference benchmarks. We then evaluate \emph{Samba-CoE} on the SN40L Node and compare with the DGX A100 and DGX H100.

\subsection{Impact of Operator Fusion}
\subsubsection{Benchmarks and Setup}

\begin{table}
\centering
\def\arraystretch{1.5}%
\begin{tabular}{|p{0.9in} |p{0.4in}|p{0.4in}|p{1in}|} \hline
\textbf{Model} & \textbf{Size} & \textbf{Sequence Length} & \textbf{Configurations} \\ \hline
llama2~\cite{llama2}     & 7B & 4K & prefill, decode, train  \\ \hline
sparseGPT~\cite{s2d} & 13B & 2K & train (87.5\% sparse) \\ \hline
llama2~\cite{llama2} & 70B & 4K & prefill, decode \\ \hline
bloom~\cite{bloom} & 176B & 8K & prefill, decode  \\ \hline
mistral~\cite{mistral} & 7B & 2K, 4K & prefill, decode  \\ \hline
falcon~\cite{falcon} & 40B & 2K & prefill, decode  \\ \hline
llava1.5-multimodal~\cite{llava} & 7B & 4K & prefill, decode \\ \hline
FlashFFTConv~\cite{flashfftconv} & N/A & 1M & FFT Convolution for long sequence models \\ \hline
\end{tabular}
\caption{Benchmarks and their descriptions. Here, `prefill' = First token generation, `decode' = Autoregressive decoding token generation with KV cache, `train' = training.}
\label{tab:benchmarks}
\end{table}

Table~\ref{tab:benchmarks} describes the set of language model benchmarks used to quantify the impact of operator fusion. The benchmarks consist of several training and inference workloads of varying sizes. Among the inference workloads, we separate out the ``prefill'' phase -- generating the first token -- from the ``decoding'' phase -- generating the second and subsequent tokens via auto-regressive decoding. Autoregressive decoding steps take advantage of the KV cache~\cite{kvcache}, and has much lower compute and operational intensity compared to the prefill phase which constructs the KV cache. The compute graph structure of \emph{FlashFFTConv}~\cite{flashfftconv} is a complex version of the motivating example described in Section~\ref{sec:fusion}.

All experiments other than \emph{FlashFFTConv} are evaluated on a system containing eight SN40L sockets and one host, witht the exception of \emph{FlashFFTConv}, which is a smaller kernel that we evaluate on a single SN40L. We measure and compare the performance of each benchmark in three configurations:
\begin{itemize}
    \item \textbf{Unfused}: Every PyTorch operator in the model is executed as one or more kernels on the SN40L, with intermediate results materialized to DDR or HBM. 
    Kernels are scheduled for execution by software. Each kernel is still parallelized to run efficiently on the SN40L.
    \item \textbf{Fused + Software Orchestrated (SO)}: Operators are fused into fewer kernels using a combination of automatic compiler optimizations and programmer hints. Kernel scheduling is performed from host software.
    \item \textbf{Fused + Hardware Orchestrated (HO)}: Same fused kernels as above, but kernel scheduling is offloaded to hardware using the feature described in Section~\ref{subsec:agcu}.
\end{itemize}

\subsubsection{Benchmarking Results}
\begin{figure*}
    \centering
    \includegraphics[width=0.75\linewidth]{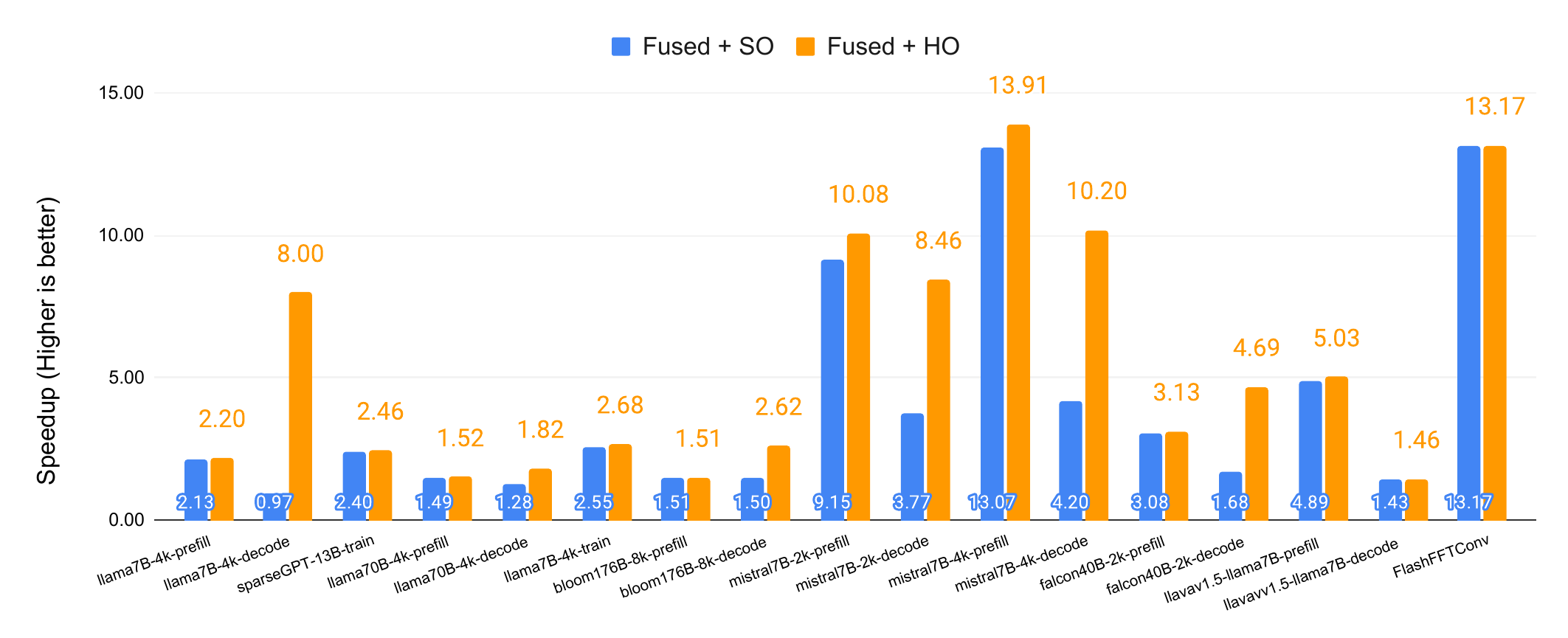}
    \caption{Measured benchmark speedups over an unfused baseline running on 8 SN40L sockets. SO = Software-Orchestrated, HO = Hardware-Orchestrated.}
    \label{fig:unfused_vs_fused_so}
\end{figure*}

\begin{figure*}
    \centering
    \includegraphics[width=0.75\linewidth]{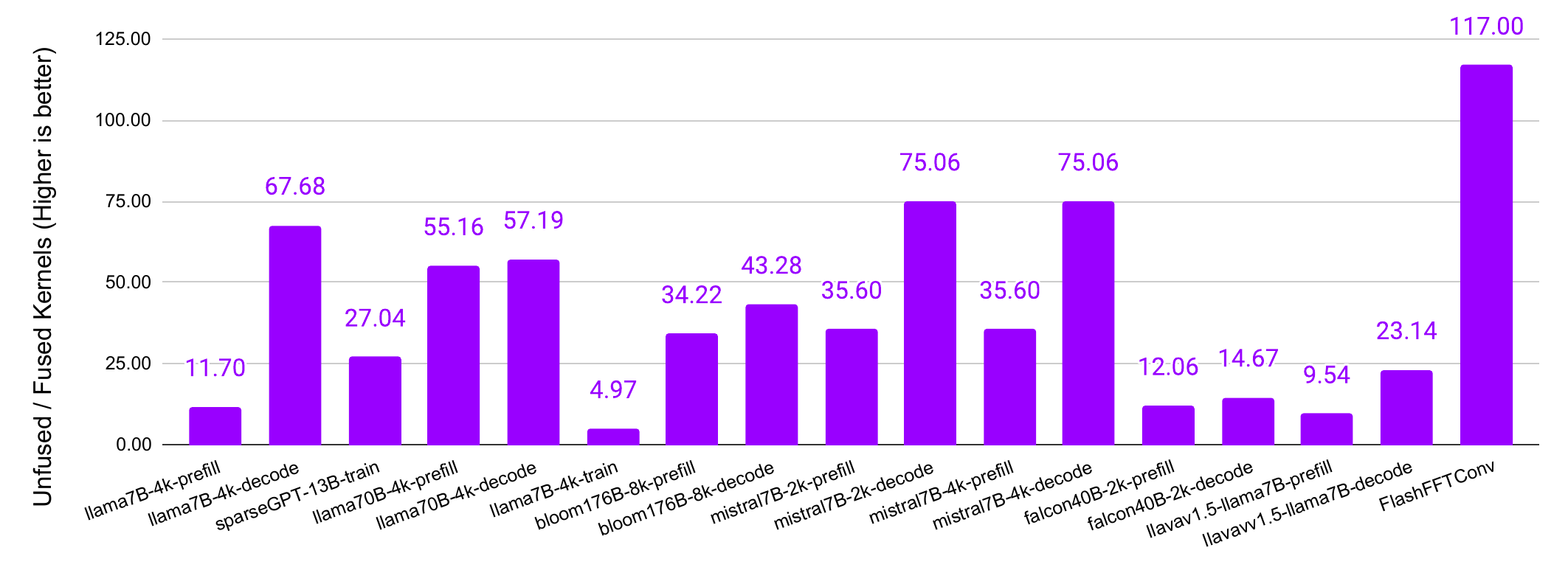}
    \caption{Ratio of number of kernel calls in unfused vs. fused configurations. 
    }
    \label{fig:kernel_call_ratio}
\end{figure*}

Figure~\ref{fig:unfused_vs_fused_so} shows the impact of operator fusion on all the benchmarks. The blue bars quantify the impact of operator fusion. The orange bars quantify the impact of hardware-orchestrated kernel launches. 

\noindent \textbf{Operator Fusion Speedup}
Without fusion, \emph{FlashFFTConv} has very low operational intensity and suffers from low performance. With fusion on the SN40L, the entire \emph{FlashFFTConv} benchmark is executed with a single kernel launch, in a manner similar to the simplified diagram shown in ~\ref{fig:motivating-df}. The increased operation intensity provides a speedup of $13\times$ over the unfused baseline.  

Inference prefill, training, and sparseGPT benchmarks achieve speedups in the $1.5\times$ to $3\times$ range. With spatial fusion, higher operation intensity is achieved with sufficient coarse-grained pipeline parallelism to keep the on-chip units occupied. For these benchmarks, batch size and sequence size provide the required level of pipeline parallelism to exploit the benefits of fusion.

In spite of having low operational intensity, autoregressive decoding inference benchmarks gain from fusion by eliminating the overheads of kernel launch and unnecessary HBM traffic. We observe speedups from $1\times$ to $13\times$, with Mistral achieving the highest speedup.

Figure~\ref{fig:kernel_call_ratio} compares the number of kernel call launches involved in executing various benchmarks in fused and unfused configurations. A ratio of 11$\times$ for \emph{llama7B-4k-inf-prefill}, for instance, means that with operator fusion, \emph{llama7B-4k-inf-prefill} was executed with 11$\times$ fewer kernel launches than its unfused baseline. This ratio quantifies the level of fusion performed in a given benchmark. For some benchmarks, higher numbers imply more aggressive fusion, like in the case of \emph{FlashFFTConv} and \emph{sparseGPT}. For others like \emph{llama70B}, higher numbers are also due to the model's larger size. 

\noindent \textbf{Hardware-Orchestrated Kernel Launch Speedup:} We now discuss the improvements obtained due to hardware-orchestrated kernel launches. Here, we see the opposite trend to the previous discussion: Autoregressive decoding inference benchmarks achieve a noticeable speedup of 1.4$\times$ to 8$\times$. Kernels have very short execution times in these benchmarks, most of which is dominated by loading weights and other inputs. Consequently, kernel launch overheads start to account for a larger fraction of the total time. Offloading kernel scheduling to the SN40L cuts out the overheads of software scheduling to provide a speedup.

On the other hand, inference prefill and training benchmarks only see a maximum improvement of  1.1$\times$. Each kernel executes for much longer in these benchmarks, and hence kernel launch overheads are amortized. The fused \emph{FlashFFTConv} benchmark has just a single kernel call, which takes the same duration with both kernel scheduling methods.

\subsection{Llama3.1 on SN40L}
\label{subsec:llama31}
We demonstrate the impact of operator fusion on the performance of Llama 3.1~\cite{llama-herd} on the SN40L. At the time of writing, Llama 3.1 is the most powerful open source model in the world with three variations - 8B, 70B, and 405B. 

\begin{table}
\centering
\def\arraystretch{1.5}%
\begin{tabular}{|p{1.2in} |p{1in}|} \hline
\textbf{Model} & \textbf{Output tokens/second/user}     \\ \hline
Llama 3.1 Instruct 405B &  $129$  \\ \hline
Llama 3.1 Instruct 70B  &  $457$  \\ \hline
Llama 3.1 Instruct 8B  &  $1042$ \\ \hline
\end{tabular}
\caption{Output tokens per second per user for Llama3.1~\cite{llama-herd} models measured on 16 SN40L sockets. All variations use BF16 weights, and mixed BF16/FP32 activations. Sequence length of 8K is used.}
\label{tab:llama31-speeds}
\end{table}

Table~\ref{tab:llama31-speeds} shows the token generation speeds for all Llama3.1 variants on 16 SN40L sockets. Speculative decoding~\cite{spec-decode} is employed on 70B and 405B models. As the SN40L fuses entire decoders into a single kernel call, almost all overheads in decoding is eliminated. Fusion enables using HBM bandwidth only for stream weights and KV cache values. Dataflow enables overlapping weight loads with computatation to achieve over 85\% of HBM bandwidth. In contrast, state-of-the-art optimized GPU implementations on H100 rarely exceed 50\% HBM bandwidth usage on weights and KV caches due to other inefficiencies. At the time of writing, the SN40L is the fastest platform in the world for Llama3.1 405B inference. This is despite the fact that inference on SN40L is performed in full BF16 precision, while other optimized GPU implementations quantize weights to 8-bit formats~\cite{aa}.

\subsection{Composition of Experts}
\label{subsec:coe}
We quantify the latency and system footprint of deploying Samba-CoE with increasing expert counts on the SN40L Node vs. DGX A100 and DGX H100. Table~\ref{tab:coe-speedup} summarizes the results. We study two scenarios with increasing expert counts: \textbf{Latency Impact} on a single node, and \textbf{System Footprint Impact} to sustain the same latency.

\begin{table}
\centering
\def\arraystretch{1.5}%
\begin{tabular}{|p{2in} |p{0.5in}|p{0.5in}|} \hline
\textbf{Metric} & \textbf{vs. DGX A100} & \textbf{vs. DGX H100}     \\ \hline
Overall Speedup, BS = 8, 20 output tokens &  $6.6\times$ & $3.7\times$        \\ \hline
Overall Speedup, BS = 1, 20 output tokens  &  $4.8\times$ & $2.8\times$     \\ \hline
Expert Speedup, BS=1, 20 output tokens  &  $2.0\times$ & $1.5\times$     \\ \hline
Overall Speedup, BS = 8, 200 output tokens &  $4.2\times$ & $2.7\times$        \\ \hline
Overall Speedup, BS = 1, 200 output tokens  &  $3.9\times$ & $2.6\times$     \\ \hline
Expert Speedup, BS=1, 200 output tokens  &  $3.2\times$ & $2.3\times$     \\ \hline
Model Switching Time                   &  $31\times$  & $15\times$       \\ \hline
$>$ 150 Experts                       &  DGX OOM     & DGX OOM       \\ \hline
\end{tabular}
\caption{Samba-CoE Performance Comparison between SN40L Node, DGX A100, and DGX H100.}
\label{tab:coe-speedup}
\end{table}

\noindent \textbf{Latency Impact:} We model two use cases: a chatbot conversation use case to produce 20 output tokens 
per input prompt, and a translation use case to produce 200 output tokens~\cite{nvda-llama2-nemo}.

The models and router are mapped as tensor-parallel over eight sockets (TP8) fashion on all platforms. The router and KV cache is always in HBM. The SN40L Node numbers are measured on real hardware. As Samba-CoE is not deployed on DGX, we estimate latencies using published model latency numbers~\cite{nvda-llama2-nemo} and optimistic model switching estimates based on DGX specs~\cite{dgx-a100, dgx-h100, a100-gpu, h100-gpu}. The total latency includes the time to run the router, copy the required expert weights, and running the expert. Two expert scenarios are studied: generating 20 output tokens, and generating 200 output tokens. We optimistically assume on the DGX that the entire capacity of HBM and host memory is available to hold weights and the KV-cache (HBM-only).

\begin{figure}
     \begin{subfigure}[b]{0.48\textwidth}
        \centering
         \includegraphics[width=0.9\linewidth]{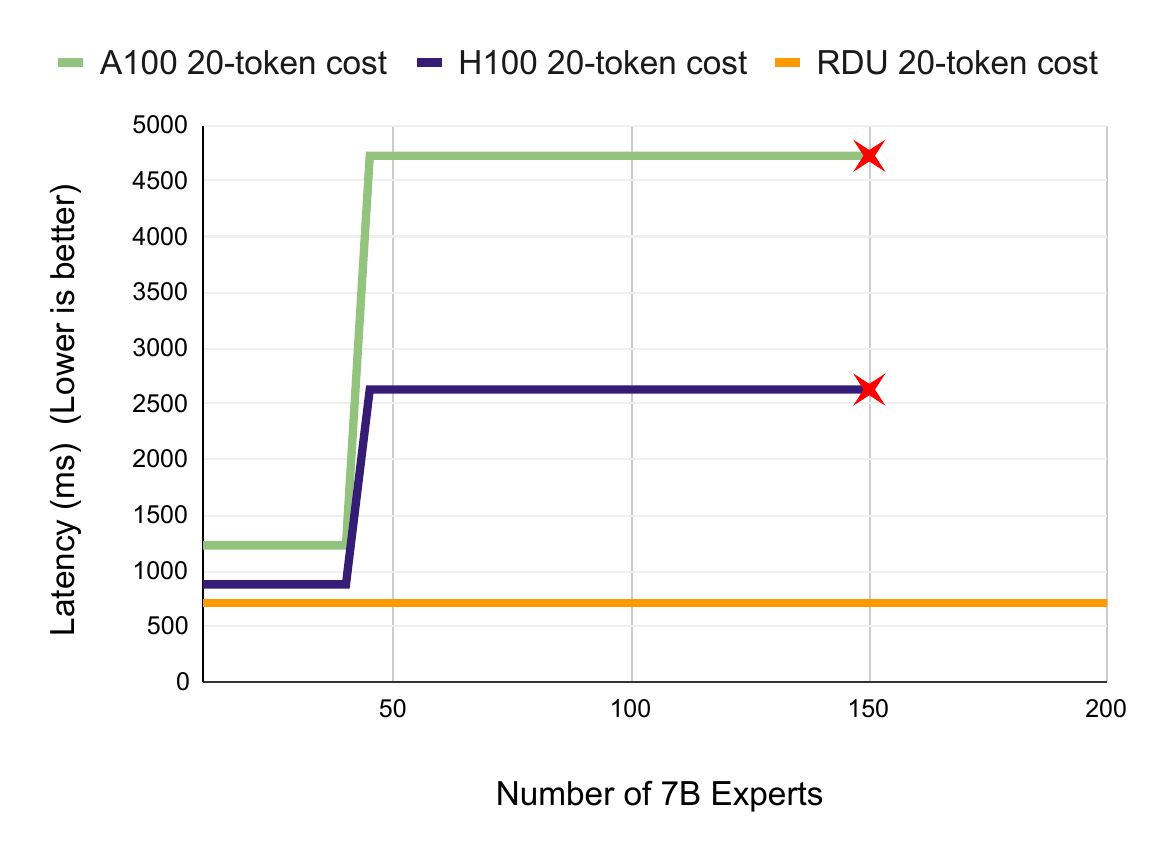}
         \caption{BS=8, TP=8 Latency.}
         \label{fig:bs8-tp8-latency}
     \end{subfigure}
     \begin{subfigure}[b]{0.48\textwidth}
        \centering
         \includegraphics[width=0.9\linewidth]{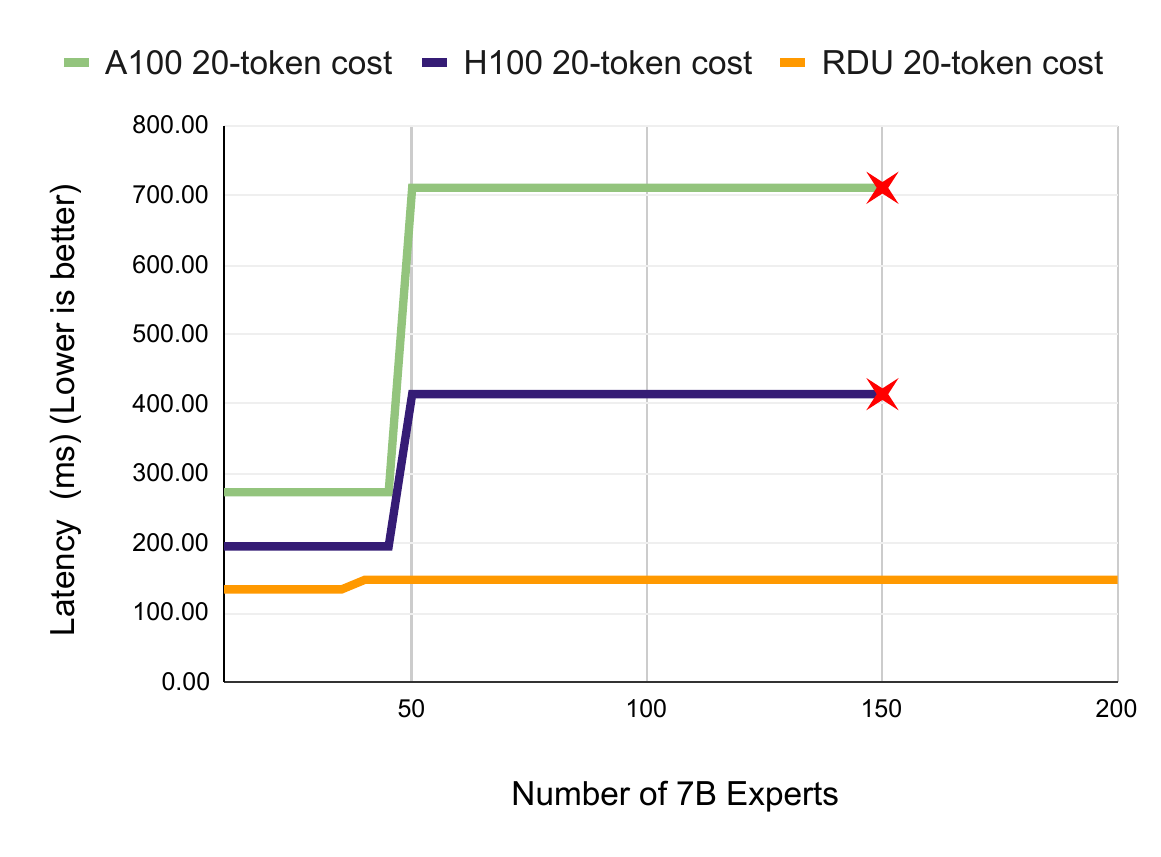}
         \caption{BS=1, TP=8 Latency.}
         \label{fig:bs1-tp8-latency}
     \end{subfigure}
        \caption{Samba-CoE latency comparison to generate 20 tokens with batch size=1 on the SN40L Node, DGX A100, and DGX H100. Batch size=8 and 200 token output configurations follows a similar trend, with speedups reported in Table~\ref{tab:coe-speedup}.
        }
        \label{fig:latency}
\end{figure}

We report latencies for batch size (BS) = 1 and BS=8 cases.
Note that ``batch'' applies to the Samba-CoE model as a whole, and not to individual experts. 
BS=8 implies that the Samba-CoE model received 8 prompts in a batch. The router is first run with BS=8 to obtain the expert for each prompt. Each prompt in the sample could need a different expert, as samples in a batch have no relationship with each other. The required experts are then copied over to HBM. Each (prompt, expert) pair is then run sequentially. 

Figure~\ref{fig:latency} compares latencies across the three platforms. We analyze these results in two broad categories:

\noindent \textbf{Under 50 experts:} All experts fit in HBM, so performance is limited by the expert execution time. The SN40L Node is $2\times$ faster than the DGX A100 and $1.5\times$ faster than the DGX H100 to generate 20 tokens. For 200 tokens, the speedup numbers are $3.2\times$ and $2.3\times$, respectively. Note that generative inference is memory-bound, and the SN40L Node has comparable HBM bandwidth to A100 (and lower than H100). The speedups clearly demonstrate the benefits of streaming dataflow: the \emph{entire decoder} layer is fused into a single kernel call, using almost 90\% of the PCUs and PMUs, and saturating close to 85\% of HBM bandwidth. Furthermore, as the model mostly contains multiple identical decoder layers, SN40L sees virtually zero kernel launch overheads.

\noindent \textbf{Over 50 experts:} Latency spikes on DGXs (around 50 7B experts) happens when experts spill over to host DRAM. For BS=8, the SN40L Node achieves a speedup of $6.6\times$ and $3.7\times$ over DGX A100 and DGX H100, respectively. For BS=1, the SN40L Node achieves speedups of $4.8\times$ and $2.8\times$ respectively. BS=8 requires copying a larger number of experts, and hence accounts for a larger fraction of the total time. Figures~\ref{fig:bs8-breakdown} and~\ref{fig:bs1-breakdown} show the time breakdown for expert switching vs. model execution. With over 1 TB/s of aggregate DDR-to-HBM bandwidth, the copy time on the SN40L Node is $31\times$ faster than DGX A100 (which provides 32 GB/s host-to-GPU bandwidth~\cite{a100-gpu}), and $16\times$ faster than H100 (which provides 64 GB/s host-to-GPU bandwidth~\cite{h100-gpu}). DGXs run out of memory at 150 experts.

\noindent \textbf{System Footprint Impact:} Next, we quantify the system footprint of increasing experts to sustain the same TP8 latency on each platform. Achieving this requires eliminating expert copies on the GPU. Consequently, all experts should reside in GPU HBM. Switching cost on the SN40L DDR to HBM is factored into the latency for SN40L.

\begin{figure}
    \centering
    \includegraphics[width=0.85\linewidth]{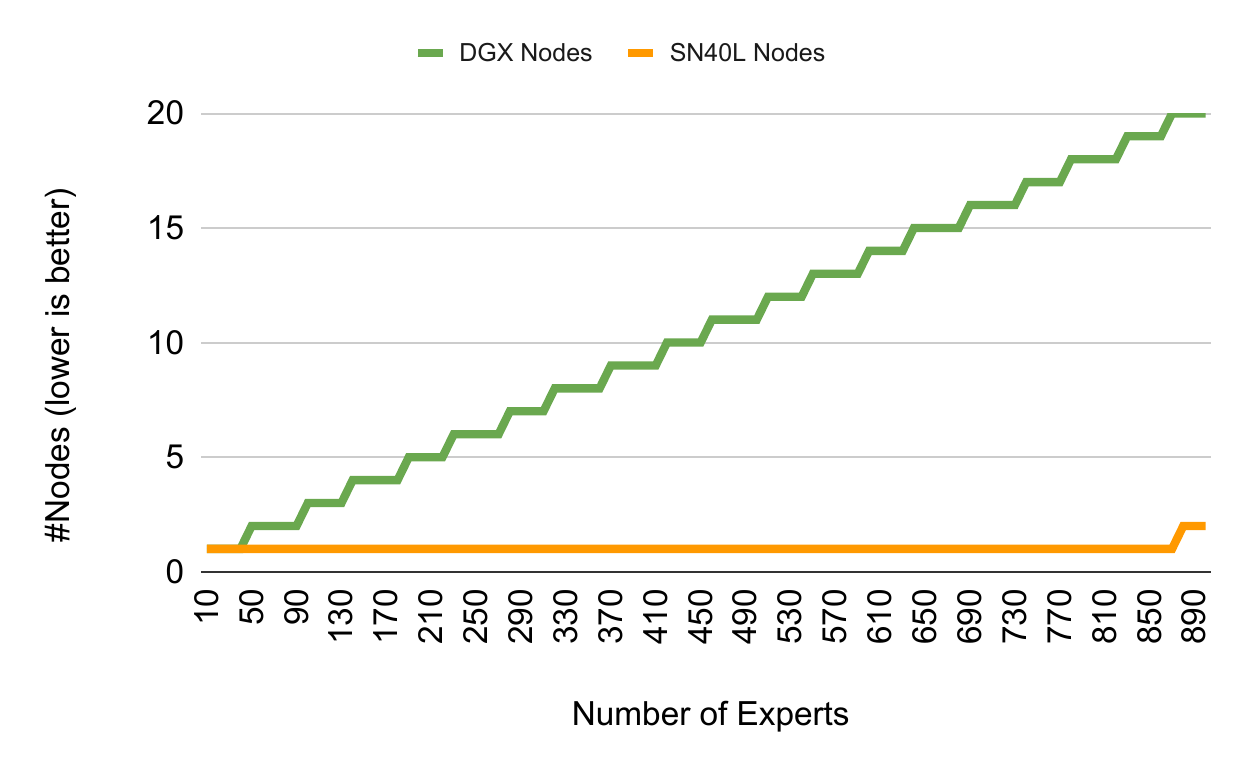}
    \caption{System footprint to sustain TP8 performance with increasing expert counts.
    }
    \label{fig:system-footprint}
\end{figure}

A single SN40L Node can hold and serve a CoE of up to 850 experts at the TP8 latency. Achieving this with DGX would need 19 DGX nodes to hold all experts in HBM.
\section{Lessons Learned}
\label{sec:discussion}
The SN40L is a complex system that is a product of the collective wisdom and hard work of many software and hardware engineers over multiple years. In this section, we discuss three lessons learned along the way.

\noindent \textbf{Managing bandwidth in software}: Software must manage bandwidth from various entities: Tile-level unit communication, HBM, DDR, die-to-die, peer-to-peer, and host bandwidth. Bandwidth translates to one or more concurrent data streams that flow in the TLN and RDN. To utilize more bandwidth from units like HBM, more load and store data streams need to be created by software. Conversely, units needing less bandwidth should be allocated fewer resources to avoid over-provisioning and wastage. Building a static bandwidth model in the compiler to model both application requirements and hardware characteristics was essential to enable proper bandwidth allocation and traffic management. The investment into a good static bandwidth model paid off in other ways: applications can be analyzed and tuned for performance to a first order statically.

\noindent \textbf{Performance debugging}: Aggressive operator fusion and pipelining creates a lot of concurrent on-chip traffic streams which create bandwidth bottlenecks. We noticed that bandwidth issues often boiled down to one of two things: a network congestion, or a memory bank conflict. Performance counters in SN40L switches and PMUs count stalls and help identify hotspots in the SN40L tile. On RDN congestion, we observe that bursty traffic can easily slow down the entire kernel if left unmanaged. Programmable packet throttling capabilities in hardware enables software to reduce bursty behavior and mitigate many RDN congestion issues. To handle PMU bank conflicts, we observed that PMUs are often programmed as double buffers of arbitrary tensor shapes, and bank conflicts could be avoided if these buffers were statically mapped to different banks.
Programmable bank bits described in Section~\ref{subsec:PMU} helped act upon this insight and eliminate bank conflicts for such multi-buffer configurations.

\noindent \textbf{Pipelined Compute and Collective Communication:} To the higher levels of our compiler, the problem of mapping data/tensor/pipeline parallel dataflows across sockets is similar to the problem of mapping them within a socket. With the hardware peer-to-peer protocol described in Sec.~\ref{subsec:agcu}, collective communication operators can be fused and pipelined with other computations into the same kernel, just like any other group of operators. Furthermore, the streaming peer-to-peer protocol between SN40Ls avoided a hop to HBM in many cases which helped conserve some communication resources.

\section{Related Work}
\label{sec:related_work}
\subsubsection{Commercial AI Accelerators}
In this section, we broadly discuss other commercial AI acceleration systems. As this is a competitive market segment, many details about contemporary AI hardware is often unknown or obscured.

Architectures like the NVIDIA A100~\cite{a100}, H100~\cite{h100}, AMD MI300X~\cite{mi300x}, Google TPUv4~\cite{tpuv4, tpuv4i}, and Intel Gaudi~\cite{gaudi} are all AI accelerators with HBM. While these architectures differ widely in their programming model, memory system, and scale-up capabilities, they do not have a three-tier memory system required to execute large CoEs and huge models efficiently. Consequently, deploying CoE on them incurs the inefficiencies described in Section~\ref{subsec:coe}. Furthermore, the streaming dataflow in SN40L provides a unique differentiation over the above as quantified in Section~\ref{sec:case_studies}. Prior studies quantify and exploit operator fusion on TPUs~\cite{fusion_asplos22} and GPUs~\cite{flashattention, flashattention2}, but do not perform the level of aggressive fusion described in this paper. Finally, the SN40L provides about $2.5\times$ higher aggregate memory capacity per socket over the recently announced NVIDIA GH200~\cite{gh200}, which enables supporting much larger CoEs models on the SN40L.

Companies like Graphcore~\cite{graphcore}, Cerebras~\cite{cerebras}, Groq~\cite{groq}, and earlier generations of SambaNova's RDU~\cite{sn10_hc, sn10_isscc} offer alternate AI accelerators. However, they all lack the three-tier memory system required to execute CoEs efficiently. To the best of our knowledge, the SN40L is the only system that has demonstrated successfully deploying a trillion-parameter CoE and other huge models in a single node and achieve the reported performance.

\subsubsection{Research Dataflow Architectures}
The topic of dataflow architectures has several prior publications covering various aspects of compute, memory, interconnect, and programming models, as covered in survey papers~\cite{cgra_survey1, cgra_survey2}. To the best of our knowledge, SN40L is the first dataflow architecture that combines streaming dataflow with a three-tier memory system, and quantify its impact on real world benchmarks.

\subsubsection{Operator Fusion}
Conventional operator fusion is a well-studied topic~\cite{fusion_asplos22, fusion-bert, pytorch2, fusion-tensorrt}. However, streaming dataflow pipelines mapped on PCUs and PMUs commonly contain 20+ operators (see Figure~\ref{fig:kernel_call_ratio}) that are automatically generated from the Python framework level by the compiler. In contrast, conventional operator fusion targets 1-5 operators~\cite{fusion-tensorrt, fusion-bert} that are often handwritten~\cite{flashattention, flashattention2, flashfftconv}, and with access pattern restrictions~\cite{pytorch2}.

\subsubsection{Parameter-efficient Fine-tuning Techniques (PEFT)}
Techniques like LoRA~\cite{lora} are commonly employed to shrink expert weights to small, low rank adapters applied to a base model. However, PEFT techniques do not achieve the same level of quality as Supervised Fine-Tuning (SFT) under several scenarios~\cite{urmish_peft_1, urmish_peft_2, urmish_peft_3, peft4, peft5}. Consequently, the smaller expert models are often entire models that are specialized using additional training or SFT (there are over 9000 variants of Llama 2 on HuggingFace at the time of this writing).

\section{Conclusion}
\label{sec:conclusion}
In this paper, we described \textit{Composition of Experts (CoE)} as a modular and cost-efficient alternative to large monolithic LLMs. 
We described the Samba-CoE with 150 experts, and motivated hardware requirements for CoE. We then introduced the SN40L dataflow accelerator and the SN40L Node that is designed to solve the memory wall using streaming dataflow and a novel three-tier memory system.
SN40L's memory system consists of on-chip distributed SRAM, off-chip HBM, and high capacity DDR DRAM.
We discussed the software impact of managing address spaces across DDR and HBM, along with runtime complexities in deploying Samba-CoE on SN40L.
We demonstrated that streaming dataflow provides a benefit of 2$\times$ to 13$\times$ over an unfused baseline. 
We showed that deploying Samba-CoE on the SN40L Node reduces machine footprint by up to
$19\times$, speeds up expert copy time by $15\times$ to $31\times$, and achieves an overall speedup of $3.7\times$ to $6.6\times$ over DGX H100 and DGX A100, respectively.

\section*{Acknowledgment}
We thank all hardware and software engineers at SambaNova Systems who worked on the SN40L RDU. Their tireless hard work and engineering creativity helped overcome numerous obstacles and enabled bringing up the SN40L RDU hardware and software stack in record time.

\bibliographystyle{IEEEtran}
\bibliography{refs}

\end{document}